\documentclass[11pt]{article}
\usepackage{amsmath, amssymb}
\usepackage{amsbsy}
\usepackage{undertilde}

\usepackage{hyperref}

\def\bdoc{\begin{document}}
\def\edoc{\end{document}}

\def\beq{\begin{equation}}
\def\eeq{\end{equation}}
\def\bea{\begin{eqnarray}}
\def\eea{\end{eqnarray}}
\def\ben{\begin{enumerate}}
\def\een{\end{enumerate}}
\def\nn{\nonumber}
\def\la{\langle}
\def\ra{\rangle}
\def\a{\alpha}
\def\b{\beta}
\def\g{\gamma}\def\G{\Gamma}
\def\d{\delta}\def\D{\Delta}
\def\e{\epsilon}

\def\th{\theta}
\def\k{\kappa}
\def\l{\lambda}
\def\m{\mu}
\def\n{\nu}
\def\o{\omega}
\def\p{\pi}
\def\r{\rho}
\def\s{\sigma}
\def\t{\tau}
\def\L{{\cal L}}
\def\S{\Sigma }
\def\gsim{\; \raisebox{-.8ex}{$\stackrel{\textstyle >}{\sim}$}\;}
\def\lsim{\; \raisebox{-.8ex}{$\stackrel{\textstyle <}{\sim}$}\;}
\def\gtrsim{\gsim}
\def\lessim{\lsim}
\def\loc{{\rm local}}
\def\vm{v_{\rm max}}
\def\bh{\bar{h}}
\def\del{\partial}
\def\nab{\nabla}
\def\half{{\textstyle{\frac{1}{2}}}}
\def\fourth{{\textstyle{\frac{1}{4}}}}
\def\third{{\textstyle{\frac{1}{3}}}}

\def\bA{{\bf A}}
\def\bD{{\bf D}}
\def\bH{{\bf H}}
\def\bM{{\bf M}}
\def\bN{{\bf N}}
\def\bE{{\bf E}}
\def\bF{{\bf F}}
\def\bB{{\bf B}}
\def\bP{{\bf P}}
\def\bJ{{\bf J}}
\def\bK{{\bf K}}
\def\bL{{\bf L}}
\def\bR{{\bf R}}
\def\bS{{\bf S}}
\def\bV{{\bf v}}
\def\bv{{\bf v}}
\def\bx{{\bf x}}
\def\by{{\bf y}}
\def\bz{{\bf z}}
\def\ba{{\bf a}}
\def\bd{{\bf d}}
\def\bs{{\bf s}}
\def\bn{{\bf n}}
\def\bm{{\bf m}}
\def\bp{{\bf p}}
\def\bk{{\bf k}}
\def\bg{{\bf g}}

\def\br{{\bf r}}
\def\bnab{{\bf \nab}}

\def\bitP{\boldsymbol{P}}
\def\bphi{\boldsymbol{\phi}}
\def\btau{\boldsymbol{\tau}}
\def\bo{\boldsymbol{\omega}}
\def\bO{\boldsymbol{\Omega}}
\def\O{\Omega}

\def\tb{\tilde{\b}}

\def\qd{\dot{q}}

\def\gt{\widetilde{g}}
\def\ft{\utilde{f}}
\def\cL{{\cal L}}

\bdoc

\begin{center} {\Large \bf
Gravitation and vacuum entanglement entropy}\footnote{\it Essay 
written for the Gravity Research Foundation 2012 Awards for Essays on Gravitation;
submitted March 31, 2012.}
\end{center}

\vskip 5mm
\begin{center} \large
{{Ted Jacobson
}}
\end{center}

\vskip  0.5 cm
{\centerline{\it Maryland Center for Fundamental Physics}}
{\centerline{\it Department of Physics, University of Maryland}}
{\centerline{\it College Park, MD 20742-4111, USA}
{\centerline{\it jacobson@umd.edu}}

\vskip 1cm

\begin{abstract}
The vacuum of quantum fields contains correlated fluctuations.
When restricted to one side of a surface these
have a huge entropy of entanglement that scales with 
the surface area. If UV physics renders this entropy
finite, then a thermodynamic argument implies the existence of
gravity. That is, 
the causal structure of spacetime must be dynamical
and governed by the Einstein equation with Newton's
constant inversely proportional to the entropy density. Conversely,
the existence of gravity makes the entanglement entropy
finite. This thermodynamic reasoning is powerful despite 
the lack of a detailed description of the dynamics
at the cutoff scale, but it has its limitations. In particular,
we should not expect to understand 
corrections to Einstein gravity in this way.

\end{abstract}

\newpage
In the vacuum state of a quantum field, fluctuations 
are correlated at spacelike separations. This entanglement  
implies that, although the vacuum is a pure quantum state, 
its restriction to a localized region is mixed. The corresponding
entropy is dominated by the shortest wavelength modes, 
and scales as the area of the region boundary\cite{Sorkin83, Bombelli86, Srednicki93,Solodukhin:2011gn}. 
In an ultraviolet (UV)
complete relativistic quantum field theory on a {\it fixed} spacetime
the entanglement entropy is infinite. 

More specifically, 
in Minkowski spacetime, the vacuum of a relativistic quantum
field restricted to the wedge
$z>|t|$,  lying to one side of an infinite $xy$ plane, is a thermal state
at temperature $T=\hbar/2\pi$ with respect to the Hamiltonian that 
generates Lorentz boosts (hyperbolic rotations) normal to the 
plane\cite{Bisognano75, Bisognano76, Unruh84}.\footnote{I adopt 
units with the speed of light 
equal to one, and assume for concreteness
that the spacetime is four dimensional. The considerations discussed here
apply in any spacetime dimension.} 
(An observer localized on a particular worldline with uniform acceleration $a$
has proper time equal to the hyperbolic angle divided by $a$, so the state is
thermal at the Unruh temperature $T_U=\hbar a/2\pi$ with respect to
the Hamiltonian generating his proper time translations.)  
The entropy is infinite
on account of the arbitrarily short wavelength fluctuations close to the 
horizon $z=t=0$, which are entangled with partners similarly close on the
other side of the horizon. If the contributions are 
cut off at a length $\ell_c$, one
obtains an entropy that scales as $A/\ell_c^2$, the area of the 
plane in units of the cutoff length.
There is nothing inherently sick in the notion of 
infinite horizon entropy;
on the contrary, what is puzzling is how 
horizon entropy could ever be {\it finite}. 

Remarkably, however, the assumption that horizon entanglement
entropy is somehow rendered finite by UV physics implies that the 
spacetime causal structure is dynamical, and that the metric satisfies 
Einstein's equation as a thermodynamic equation of state. 
That inference arises as follows\cite{Jacobson95}. 
Suppose that the entropy area density 
of any local causal horizon 
is $\a<\infty$, and that the entropy satisfies the usual thermodynamic
Clausius relation
\beq\label{Clausius}
\d S = \d Q/T
\eeq
for all such horizons, with $\d Q$ the (approximately defined) boost 
energy flux and $T=\hbar/2\pi$ the boost temperature of the vacuum
mentioned above.
Then the spacetime geometry {\it cannot be inert}: the light
rays generating the horizon must focus so that
the area responds to the flux of energy in just the
way implied by the Einstein equation (at least at long distances).
The cosmological constant is undetermined, and 
the value of Newton's constant given by $G=1/4\hbar\a$
{\it which is nonzero provided the entropy density $\a$ is finite}.
This implies that the
entropy density $\a=1/4\hbar G =1/4L_p^2$
is $1/4$ in units of the Planck length $L_p=\sqrt{\hbar G/c^3}$, 
in agreement with the Bekenstein-Hawking entropy 
\beq\label{SBH}
S_{\rm BH}=A/4L_p^2
\eeq
inferred long ago for
black hole horizons\cite{Bekenstein73,Hawking74}, 
cosmological horizons\cite{Gibbons77}, 
and acceleration horizons\cite{Hawking95, Massar99}.

But how could the horizon entropy ever be finite? A naive 
cutoff at some short distance would select a 
preferred reference frame, in violation of Lorentz symmetry. Since
Lorentz boost symmetry lends the vacuum its thermal
character, this seems a rather unlikely means to regulate the entropy.
Moreover, Lorentz violation appears to wreak havoc with black hole
thermodynamics, as it entails violations of the generalized
second law\cite{Dubovsky:2006vk,Eling:2007qd,Jacobson:2008yc}. 
Thus we should look elsewhere to understand
the finiteness of horizon entropy. 

Since the entropy can certainly be infinite in a theory 
with no gravity, and since the finiteness assumption implies 
gravity, it seems that gravity itself should somehow
render the entropy finite.
A natural idea, proposed before by many researchers\cite{Solodukhin:2011gn},
is that quantum gravitational fluctuations of spacetime, which 
are expected to be large at the
Planck scale $L_p$,  somehow cut off the entropy of UV modes 
approaching the horizon at shorter distances. 
This yields an entropy density of order $L_p^{-2}$, matching the
Bekenstein-Hawking entropy. We can be a little more specific about
how this might work.
It would violate translation invariance
of the vacuum if there were any particular location at which 
spacetime fluctuations became large, and it would violate
Lorentz invariance if there were any particular length scale at which they
became large. Rather, they are large
everywhere when considered in some invariant sense. 
The invariant that is relevant for vacuum entanglement is the 
proper separation between the correlated fluctuations.
 
The gravitationally dressed ground state satisfies the quantum analog of the
initial value constraints of general relativity, the Wheeler-DeWitt equation. 
This equation correlates the constrained part of the gravitational field 
to distributions of energy. 
A pair of entangled fluctuations localized and separated by a proper distance $\ell$
have an associated quantum energy $E\sim \hbar/\ell$, and so 
must entail an associated metric perturbation of order 
$GE/\ell\sim L_p^2/\ell^2$. When $\ell<L_p$ the perturbation
is large, and the causal structure of the spacetime
is strongly modified. The gravitationally dressed vacuum fluctuations
thus cannot be separated by a fixed 
horizon when they are closer than a Planck length away from each other.
In effect, the entanglement is ``cloaked".

While I have argued that gravity can make the entanglement entropy finite, 
the argument
of course does not establish that it is precisely equal to the Bekenstein-Hawking entropy.
In fact, as presented so far, it does not even establish that the entropy
is of order $A/L_p^2$. 
For one thing, it does not take into account the fact that
the entropy should (presumably) grow with the number $N$ of field species in nature. 
Also it does not account for the running of the gravitational coupling constant 
$G$ with 
the length scale. To include those effects, we must
allow for the fact that $G$ at scale $\ell$ 
is determined by its value at low energies $G_0$ and the number and nature of 
field species.
For a theory with $N$ fields of the same type
we should replace ``$G$" in the above analysis by $G(G_0, \ell, N)$.
Then 
the condition determining the cutoff length $\ell_c(G_0,N)$ becomes
\beq
\ell_c^2  = \hbar G(G_0, \ell_c, N).
\eeq
(Here
I am assuming that the energy that determines the gravitational dressing of 
a correlated pair separated at scale $\ell$ is still $\hbar/\ell$, independent of the field species.
See below for further discussion.)
If the entropy grows roughly in proportion to $N$
(aside from the $N$-dependence of $\ell_c$), 
it would then be given by
\beq
S\sim \frac{NA}{\ell_c(G_0,N)^2}. 
\eeq
This would scale as the Bekenstein-Hawking entropy (\ref{SBH})
provided
the renormalization works out such that
\beq
\label{consistency}
\frac{N}{[\ell_c(G_0,N)]^2}\sim \frac{1}{\hbar G_0}.
\eeq
It makes some sense that the reciprocal of the low energy Newton constant 
$G_0$ is proportional to $N$ in units of $G$ at the
cutoff scale, since the $N$ species would each contribute to the
low energy effective gravitational 
action\cite{Susskind:1994sm,Callan:1994py,Jacobson:1994iw}. 

However, while it looks superficially satisfactory,
this reasoning is rather incomplete.
For one thing, the theory is strongly 
coupled at the cutoff scale, and it isn't really clear how the 
number of species affects the result except perturbatively.
Also, since there are $N$ times as many
independent fluctuating fields at each scale,
one might naively expect an extra factor of $\sqrt{N}$
weighting the fluctuation energy, which would change the 
scaling of $\ell_c$ with $N$. Moreover,
nonminimally coupled fields with certain sign of the 
nonminimal term can push the flow of the gravitational coupling 
in the opposite direction\cite{Solodukhin:1995ak, Larsen:1995ax,Barvinsky:1995dp},
and gauge vector\cite{Kabat:1995eq} and graviton tensor\cite{Fursaev:1996uz} 
fields seem (at least using existing methods) 
to do the same.
And, finally, the initial condition for the
renormalization flow needs to be set. A simple relation
of the form (\ref{consistency}) makes sense only
if the entire low energy gravitational action is induced by the
matter fluctuations\cite{Jacobson:1994iw}. That could be so, but 
there is no apparent reason to suppose that it \textit{must} be so.

All these complications seem to call into question the validity
of the notion that entanglement entropy lies at the root of 
the Bekenstein-Hawking entropy. However, the thermodynamic
derivation  of the Einstein equation mentioned above
sidesteps all of these difficulties.  
According to that derivation,
if the horizon entanglement entropy is 
finite and satisfies the Clausius relation (\ref{Clausius}) 
(as should any entropy near 
equilibrium) then, whatever the underlying UV physics, 
it will always be equal to 
the Bekenstein-Hawking entropy
with respect to the low energy Newton constant that appears in 
Einstein's equation. 

This sounds rather satisfactory, but it can be questioned from another direction.
The Einstein equation is presumably just the lowest
order approximation to a field equation that has higher
curvature terms, and
in such theories the entropy of stationary
black hole horizons involves curvature corrections\cite{Iyer:1994ys,Visser:1993nu,Jacobson:1993vj}.
Hence, to obtain such corrections to the 
equation of state, we should presumably begin with corrections to the
horizon entropy function, for example constructed from the local curvature
tensor and horizon geometry. 
However, 
despite many attempts\cite{Eling:2006aw,Elizalde:2008pv,Brustein:2009hy,Parikh:2009qs,
2009arXiv0903.1254P, Chirco:2010sw,Guedens:2011dy} to show it,
except for the very special case where the entropy density 
is just a function of the spacetime 
Ricci scalar,
the Clausius relation for an intrinsic entropy of 
all local causal horizons
does not appear to be equivalent to a local tensor field equation.
(In Ref.~\cite{Guedens:2011dy} we did manage to 
consistently derive some higher derivative
field equations in this way, but the assumed entropy depended on
arbitrary features of an approximate local Killing vector, so was not 
intrinsic to the spacetime and horizon.)

Should we infer from this failure that the thermodynamic derivation is just 
a fluke that works for Einstein gravity, but is not really fundamental?
I think not. There is a good physical reason for the failure. 
The ``heat" $\d Q$ in the Clausius relation (\ref{Clausius}) 
is taken as the boost energy flux, whose definition involves 
an approximate local boost Killing vector $\xi$. A flat spacetime has exact 
boost Killing vectors, but in a curved spacetime $\xi$ is 
defined only up to ambiguities of order $(\ell/L_{\rm curv})^2$, where 
$L_{\rm curv}$ is the typical local radius of curvature of the spacetime
in some frame
and $\ell$ is the length over which $\xi$ is defined. 
Now consider what happens if there is 
a curvature term $L_1^2 R$ in the entropy density,
with $L_1$ some constant with dimensions of length and $R$ some
curvature quantity.
This is suppressed relative to the area term by a factor
$(L_1/L_{\rm curv})^2$. If the $\xi$ ambiguity is to be smaller than 
this we must restrict to a region of size $\ell<L_1$. 
If $L_1\sim L_p$, as expected in a theory in which the Planck length is the only
UV scale, this would require
the region to be smaller than the Planck length. But we have seen that
the quantum fluctuations of the metric associated with a pair at that scale are large,
which invalidates the application of the thermodynamic Clausius relation
to such a small region of a classical horizon. Thus, in a theory with only
one UV length scale, we should not expect to be able to capture corrections
to the horizon entropy beyond the area term by a local thermodynamic argument.
(Corrections can of course be captured by global considerations
involving stationary black hole configurations.)

But how about a theory
in which $L_1\gg L_p$? Then it would appear the corrections to the entropy
could be larger than the ambiguity in $\d Q/T$, so the local thermodynamic
derivation should be able to capture them.
However, it seems that in this case the limit to localization is no longer 
$L_p$ but instead is the longer scale $L_1$. For instance, in string theory, 
the low energy effective action has $L_1\sim L_s$, where $L_s\gg L_p$ is 
the string length. When length scales smaller than $L_s$ are probed,
an infinite number of higher curvature terms in the action are equally important,
and a stringy description of the degrees of freedom
is necessary. One might imagine 
the only higher curvature term in the action is the $R^2$ term, but this requires
unnatural fine tuning, and there is probably no physical theory that actually 
behaves this way. 

The simplicity of the area entanglement entropy
belies a microscopic complexity. Yet, without knowing
in detail how to identify and count the precise degrees of freedom,
thermodynamic reasoning allows us to deduce 
the universal relation between horizon entropy and Einstein gravity.
But thermodynamic reasoning has its limitations; in particular,
it seems we should not expect to understand 
corrections to Einstein gravity in this way.

\section*{Acknowledgments}
I thank R. Myers, A. Satz, and A. Wall
for helpful discussions and comments on a draft of this essay.
This research was supported in part by NSF grant PHY-0903572,
and by the Kavli Institute for Theoretical Physics through NSF grant 
PHY11-25915.

\bibliographystyle{utphys}
\bibliography{graventanglement}

\edoc